\documentstyle[11pt,aaspp4]{article}
\received{1999 Mar 2}
\accepted{1999 Aug 4}

\slugcomment{To appear in the PASP, vol 111, 1999 Nov}

\lefthead{Welsh}
\righthead{CCF Lags in AGN}

\begin{document}

\title{
On the Reliability of Cross Correlation Function Lag
Determinations in Active Galactic Nuclei}

\author{W. F. Welsh\altaffilmark{1} }
\affil{Department of Astronomy and McDonald Observatory, 
University of Texas at Austin, Austin, TX 78712 USA}
\altaffiltext{1}{e-mail: wfw@astro.as.utexas.edu}

\begin{abstract}

Many AGN exhibit a highly variable luminosity. Some AGN also show a
pronounced time delay between variations seen in their optical continuum
and in their emission lines. In effect, the emission lines are light
echoes of the continuum. This light travel--time delay provides a
characteristic radius of the region producing the emission lines. 
The cross correlation function (CCF) is the standard tool used to measure
the time lag between the continuum and line variations.
For the few well--sampled AGN, the lag is $\sim$1--100 days,
depending upon which line is used and the luminosity of the AGN.
In the best sampled AGN, NGC 5548, the H$\beta$ lag shows year--to--year
changes, ranging from $\sim$8.7 days to $\sim$22.9 days over a span of
8 years.
In this paper it is demonstrated that, in the context of AGN
variability studies, the lag estimate using the CCF is biased too low
and subject to a large variance. Thus the year--to--year changes of
the measured lag in NGC~5548 do not necessarily imply changes in the
AGN structure.  The bias and large variance are consequences of
finite duration sampling and the dominance of long timescale trends in 
the light curves, not due to noise or irregular sampling. 
Lag estimates can be substantially improved by removing low frequency
power from the light curves prior to computing the CCF.

\end{abstract}

\section{Introduction}
Active galactic nuclei (AGN) often exhibit variable luminosity.
In several highly variable AGN, the observed UV/optical emission--line
fluxes are well correlated with the continuum variations, but with a time
delay (e.g. Peterson 1988; Ulrich, Maraschi \& Urry 1997).   
In effect, the line emission is a light echo of the photoionizing
continuum.  For Seyfert 1 galaxies, the well--measured time delays
(``lags'') range from $\sim$1--80 d,
and depend on which emission line is observed (the higher ionization lines
respond more quickly to continuum variations than do lower ionization
lines; see e.g. Clavel et al. 1991; Peterson et al. 1998b).
The observed time delay gives a characteristic timescale that, under
reasonable assumptions
(i.e. the lines are responding to photoionization, the observed
continuum closely mimics the photoionizing continuum, and the
light--travel timescale is the most relevant timescale),
provides a characteristic lengthscale 
(see e.g. Peterson 1988, 1993). Thus the observed lag between
AGN continuum and emission--line light curves gives a measure of the size
of the line--producing region, i.e., the broad--line region (BLR). 
Note that these lengthscales correspond to angular scales of
microarcseconds on the sky, so ``echo mapping'' studies offer the
potential for extremely high spatial resolution studies of AGN
(e.g. the proceedings by Peterson and by Horne in Gondhalekar, Horne \&
Peterson 1994).

The Seyfert galaxy NGC 5548 has been the target of several variability
studies and has been intensely monitored for several years (e.g. 
Peterson et al. 1999).  Of particular interest are the time lag
determinations for the H$\beta$ emission line with respect to the optical
continuum. The lag (as defined by the peak of the cross correlation
function) ranges from $\sim$8.7 d to $\sim$22.9 d over a span of 8 years
(1989-1996) with an rms scatter of 4.5 d (Peterson et al. 1999).  
These variations have
been interpreted as evidence
for real structural changes in the BLR, either due to physical changes of
the ensemble of BLR clouds or to changes in the photoionizing radiation
field (e.g. Wanders \& Peterson 1996). The dynamical timescale for the BLR
($\sim \frac{BLR
\ size}{BLR \ cloud \ velocities}
  \sim \frac{c \ \times \ lag}{line \ width}$) 
is on the order of a few years, so changes in the BLR structure on this
timescale is a realistic possibility. 

However, there is considerable difficulty in determining the lag from 
AGN time series, in particular, because (i) the data are short in
duration compared to the timescales of interest and (ii) the data are
usually not equally sampled. Previous studies have investigated the issue
of noisy and poorly sampled data, e.g., see the summary in Koen (1994).
Several different methods for computing the CCF have been constructed:
the interpolated CCF (``ICCF'' --- e.g. Gaskell \& Peterson 1987); the
discrete CCF method (``DCF'' --- Edelson \& Krolik 1988); the inverse
Fourier transform of the discrete power spectrum (Scargle 1989); and the
Z--transform CCF (Alexander 1997). In addition, methods other than the CCF
have been used to measure time lags, e.g., optimal reconstruction via
minimizing $\chi^{2}$ (Press, Rybicki \& Hewitt 1992) or a parametric
approach (i.e., modeling the light curves as random walks; Koen 1994).
For comparisons of the DCF and ICCF methods, see e.g.,
Litchfield, Robson \& Hughes (1995), White \& Peterson (1994), and
Rodriguez--Pascual, Santos--Lleo \& Clavel (1989).
Simulations have shown that these methods can provide reasonably
accurate determinations of the lag under sampling and noise
conditions similar to the actual observations (e.g. Peterson et al. 1998a;
White \& Peterson 1994, Koratkar \& Gaskell 1991b). 
Hence the changing H$\beta$ lag in NGC~5548 seems quite real.

In this paper we consider two additional sources of ``error'' in the CCF
lag determinations, which to our knowledge have not been fully addressed
in the astronomical literature. The first is a bias inherent in the
definition of the CCF such that, on average, the computed CCF lag is not
the true lag. The second error is concerned with gross changes in the CCF
due to changes in the ACF of the continuum. These changes in the ACF are
not real, in that the underlying physical process generating the
continuum variations have not changed; they are simply statistical
fluctuations inherent in any finite realization of a stochastic process.

In \S 2 we define the autocorrelation function (ACF), cross correlation 
function (CCF) and transfer function and examine the relationships
between them in the AGN echo mapping context.
Several aspects of the CCF are examined in \S3, 
and at the risk of being overly pedagogical, 
we present this material in detail because they
are crucial to the interpretation and usefulness of the CCF. 
In particular, the bias inherent in the definition of the CCF 
and the error in the lag determination are highlighted.
To illustrate and quantify these ``problems'' with the CCF, simulations
tailored to match the characteristics of the H$\beta$ observations of
NGC~5548 are presented in \S4.  The simulations clearly show the bias and
large variance present in the CCF. Additional simulations are also
presented to: (1) explore a wider range in parameter space, (2) help
quantify the amount of bias and variance, and (3) illustrate how the bias
and variance can be reduced by removing low frequency trends from the
light curves. We conclude with a discussion of our results in \S5.

\section{The ACF, CCF and Transfer Function }
\subsection{The Standard Definitions of the ACF and CCF}
The auto-- and cross--correlation functions are standard tools of 
time series analysis (e.g. Jenkins \& Watts 1969; Box, Jenkins \&
Reinsel 1994; Chatfield 1996; Wall 1996).
The cross--correlation function, sometime called the serial
correlation function, quantifies the amount of similarity or correlation
between two time series as a function of the
time shift (i.e., the delay or ``lag'') between the two time series. 
The auto--correlation function (ACF) measures the similarity of a single 
time series to a delayed version of itself. 

The standard definition of the cross--correlation function of
two time series $x_{i}$ and $y_{i}$ sampled at discrete times $t_{i}$ 
($i=1,...,N$) with equal sampling ($\Delta t = t_{i+1}-t_{i}$) is:
\begin{eqnarray}
CCF(\tau_{k}) \equiv
\frac{
\frac{1}{N} \displaystyle{ \sum_{i=1}^{N-k} }
(x_{i}-\bar{x})(y_{i+k}-\bar{y}) 
}
{ 
\left[\frac{1}{N} \displaystyle{\sum_{i=1}^{N}} 
(x_{i}-\bar{x})^{2}\right]^{1/2}
\left[\frac{1}{N} \displaystyle{\sum_{i=1}^{N}}
(y_{i}-\bar{y})^{2}\right]^{1/2}
}  
\end{eqnarray}
where the lag $\tau_{k}$ is the size of the time shift: $\tau_{k} = k 
\Delta t$, $k=0,...,N-1$ and $\bar{x}$, $\bar{y}$ are the means of $x_{i}$
and $y_{i}$ (see e.g. Jenkins \& Watt (1969),  Chatfield (1996)). 
The ACF is similarly defined, with $x_{i}$ itself in place of $y_{i}$.
[NB: for negative lags, simply interchange $x$ and $y$.] 

It will be helpful to express the CCF more succinctly, and
we will use the continuous definition to do so:
\begin{eqnarray}
CCF(\tau) = \int x(t) y(t+\tau) dt \\
ACF(\tau) = \int x(t) x(t+\tau) dt
\end{eqnarray}
It should be explicitly stated that we use equations (2) \& (3) only as
shorthand representations of equation (1), as the discrete and continuous
CCF are not the same.
Also note that in this nomenclature $\bar{x}$ and $\bar{y}$ are by
definition zero and the light curves have been normalized to unity
variance.

\subsection{The transfer function $\Psi$}

AGN broad emission--line light curves are not simply delayed 
copies of the continuum light curves.  This can be understood as a
geometrical effect, as first pointed out by Blandford \& McKee (1982):
the line--emitting region extends over a large volume of space, 
so the light travel time across the BLR is significant. The integrated
line light curve thus appears as a delayed and blurred version of the
continuum light curve.

Blandford \& McKee (1982) expressed the relationship between
the line emission $L(t)$ and the continuum emission $C(t)$ as:
\begin{equation}
L(t) = \int C(t-\tau) \Psi(\tau) d\tau.
\end{equation}
The geometry and responsivity of the gas is contained in the
 ``transfer function'' $\Psi$, and equation (4) simply states that the
line light curve is equal to the continuum light curve convolved with
the transfer function. The transfer function can be thought of as a
``point spread function'', ``impulse--response function'', or as a
filter of a linear moving average process with the continuum as the driver 
(but note that in this interpretation the continuum is {\em not} an
uncorrelated white noise process).
Only if $\Psi$ is a delta function will the $L(t)$ be
an identical (but lagged) version of $C(t)$ 
[NB: we use the term ``identical'' loosely here: we mean $L(t)$ 
is identical to $C(t)$ within a scale factor and constant, allowing for
$\int \Psi d\tau \neq 1$ and a background contribution to both the line
and  continuum. We also implicitly assume $\Psi (\tau < 0) = 0$, i.e.,
$\Psi$ is purely causal.]
Recovering the transfer function is a major goal of variability
studies in AGN, as it contains information on the geometry
and kinematics of the BLR. The reader is referred to Horne (1994), 
Pijpers \& Wanders (1994), Krolik \& Done (1995), Vio, Horne \&
Wamsteker (1994), and the proceedings in Gondhalekar, Horne \& Peterson
(1994) for a discussion of the transfer function and inverting equation
(4) to solve for $\Psi$.

\subsection{The relationship between the ACF, CCF and transfer function}
From the definitions it can be shown 
(e.g. Koratkar \& Gaskell 1991a; Penston 1991; Sparke 1993) that
\begin{equation}
CCF(\tau) = \int_{-\infty}^{\infty} \Psi(\tau') ACF_{c}(\tau - \tau')
d\tau',
\end{equation}
i.e., the cross--correlation function is equal to the 
transfer function convolved with the auto--correlation function of the
continuum. In this representation it becomes clear that the  
theoretical CCF is identical to a blurred version of the transfer
function.

If the continuum light curve consists of a well--isolated sharp pulse, or
equivalently, its power spectrum is white,  then its ACF$_{c}$ is a delta
function and the CCF is then identical to $\Psi$. In this circumstance
the peak of the CCF will occur at the same lag as the peak of $\Psi$.
More generally, the ACF is a broad and even function, so the peak of
the CCF will not necessarily occur at the same lag as the peak in $\Psi$.
The lag determined from the CCF should be considered only a
characteristic time scale.

Equation (5) concisely describes the fundamental issue we address in
this paper: the determination of the lag between line and continuum
light curves depends on both the shape of the transfer function and the
ACF of the continuum. However, it is crucial to understand that eq. (5)
is valid only in the infinite duration limit --- for finite limits,
it is straightforward to show the equality is {\em not} true.

\section{ Practical Issues }
Just as with Fourier analysis, the difference between the mathematical
CCF and the experimental (discrete and finite--sampled) CCF can be large.
Thus it is useful to examine some of the details and practical issues
concerning the computation of the CCF, with emphasis on the AGN echo
mapping problem. Several of the issues discussed in this section will be
illustrated with simulations presented in \S4.
Problems concerning unequal sampling have been discussed in the literature
(see the references in the introduction) and will not be repeated here. 

\subsection { General aspects of the CCF }
\subsubsection {``self correlation''}
The individual points in the ACF and CCF are highly correlated with
themselves, i.e., neighboring points are not independent (a derivation
can be found in Jenkins \& Watts (1969)). In general, the neighboring
values in the ACF/CCF are more correlated with themselves than 
neighboring values in the original time series (Jenkins \& Watts 1969).
It is important to be aware of this ``self--correlation'' in the
interpretation of the ACF/CCF because trends in the ACF/CCF are
long--lived, e.g. it can take a surprisingly long time for the ACF/CCF to
decay from a peak. This can also lead to spurious large values of the CCF,
especially for time series whose ACFs contain intrinsically broad peaks
--- see e.g. Figs. 1--4 in Koen (1994).

\subsubsection { bias }
The CCF as defined in equation (1) has the peculiar property
that the summation in the numerator goes from $i=1$ to $N-k$, 
but the normalization is $1/N$, not $1/(N-k)$.
This normalization is chosen primarily because the variance of the CCF is
then reduced, i.e., the sample CCF is a better estimator of the true CCF
in the mean square error sense when the $\frac{1}{N}$ normalization is
used (Jenkins \& Watts 1969).
This normalization also guarantees that the CCF is always bounded by
$\pm1$, and the autocorrelation matrix is positive semi--definite
so that the ACF and power spectrum are Fourier transform pairs
--- the Wiener--Khinchin theorem (e.g. Jenkins \& Watt 1969).
However, the $\frac{1}{N}$ normalization means that eq. (1) is only an
asymptotically unbiased estimator of the correlation function, and its use
introduces a well--known bias towards zero (e.g. Kendall 1954;
Marriott \& Pope 1954; Otnes \& Enochson 1978; Chatfield 1996). 
This bias grows worse with increasing lag, and results in a
triangular--shaped reduction of the CCF and hence underestimates the lag
of the peak of the CCF.
The trade--off between the bias and variance in an estimator is a common
problem in statistics: often one must choose between precision (low
variance) and accuracy (low bias). 
For the CCF, the bias goes as $\frac{1}{N}$ while the variance goes
as $\frac{1}{\sqrt{N}}$; this is why reducing the bias has not been
treated with equal importance as reducing the variance (Tjostheim \&
Paulsen 1983).
It is argued that in most cases $N \gg 1$ and $k\ll N$, so one can
tolerate a small bias to achieve a large the reduction in variance. 
Furthermore, eq. (1) is extremely simple to compute.
For these reasons the $\frac{1}{N}$ normalization has gained predominance. 

In the case of AGN variability, however, the ACFs are intrinsically broad
and the lags of interest are usually a substantial fraction of the
duration of the observations, hence this bias can lead to a significant
underestimation of the lag. As Scargle (1989) points out, this bias can
be devastating.  One could simply replace the $\frac{1}{N}$ term with
$\frac{1}{N-k}$, but then the CCF has the very undesirable property that
it can exceed unity.  Indeed, it is likely to exceed unity because of the
self--correlation in the CCF;  the ACF at $k$=0 is forced by definition
to be exactly 1, so due to self--correlation the value of the AGN ACF at
$k$=1 will tend to be close to 1 as well, and so on for many lags. 
As shown by  Marriott \& Pope (1954) and by Kendall (1954), the bias in
the estimated ACF at lag $\tau$ depends, in general, on the value on the
{\em true} ACF at lags $\tau$ and earlier, a consequence of the strong
self--correlation. Because of this, the bias cannot be removed {\it a
priori} and the simple $\frac{1}{N-k}$ attempt at bias correction will
in general fail. 

In his method for coping with unequally sampled data, Scargle (1989)
suggests renormalizing the ACF with the ACF of the sampling pattern,
which will remove both the effects of this bias as well as the effects of
irregular sampling. Scargle notes that this correction is on
average equivalent to replacing $\frac{1}{N}$ term with $\frac{1}{N-k}$,
and so it also allows values of the CCF to exceed unity.  

We find that in practice, the $\frac{1}{N-k}$ normalization has the
side--effect of boosting noise at large lags, but even worse, the shape
of the ACF can be modified, and peaks in the CCF at $\tau \neq 0$ can be
shifted to longer lags. This is totally unacceptable for our purposes and
so this proposed renormalization is rejected. 

To the best of our knowledge, most of the effort in reducing the bias in
the CCF as given in equation (1) has been motivated by the desire to 
accurately determine the Yule--Walker filter coefficients of a stochastic
moving average (or autoregressive) process. These filter coefficients
can be determined from the ACF, and in particular, the first few lags of
the ACF. Since this usually corresponds to $k\ll N$, a ``better''
definition of the ACF has not been sought and instead bias--corrections
have been developed to treat the specific problem of determining the
filter coefficients (e.g. Kendall 1954; Tjostheim \& Paulsen 1983;
Marriott \& Pope 1954). 
In their noteworthy analysis, Sutherland, Weisskopf \& Kahn (1978)
address the question of bias  specifically for the ACF in the case of a
noisy and finite length shot--noise light curve, and in particular, they
derive a ``partially unbiased'' definition for the ACF. 
They show that in addition to the 
bias that is present in the noise--free case, there is an additional
reduction of the ACF due to noise; this comes about because the
normalization of the ACF depends on the variance of the light curve, and
in the presence of noise, the variance itself is biased too high. Thus the
value of CCF depends on the signal--to--noise ratio (S/N) of the data. The
motivation for their work was to deduce the correlation time constant
for the shot noise in Cyg X-1. They show that this can be deduced in an
unbiased fashion from the {\em ratio} of the ACF at lags  $k$=1 and $k$=2.
It is not obvious that their partially unbiased ACF, suitable for small
lags, is applicable to the CCF at large lags but this is a topic worth
pursuing.

Another possible method to reduce the bias in the ACF is to use the
``jackknife'' or Quenouille method (see e.g. Chatfield 1996):
$ACF^{\prime} = 2 ACF - \frac{1}{2}(ACF_{1}+ACF_{2})$, where
$ACF_{1}$ is the ACF of the first half of the data set and $ACF_{2}$
is the ACF of the latter half. The bias in $ACF^{\prime}$ is reduced from
order $\frac{1}{N}$ to $\frac{1}{N^{2}}$. Although this reduces the length
of the already too short AGN time series by a factor of two, it does
allow one to check the stationarity assumption between the two halves.
While we have not investigated this bias correction method, it is worth
consideration in future studies.

\subsubsection { peak or centroid? }
The peak of the CCF gives the lag where the two time series are most
highly linearly correlated. Yet the peak of the CCF need not
correspond to the peak of the transfer function $\Psi$ --- indeed, the
transfer function may not have a well--defined peak at all.  
The peak of the AGN line--continuum CCF tends to correspond to where the
line echo response is most coherent, i.e. the innermost region of the
BLR, and hence can underestimate the BLR size (e.g. Gaskell \& Sparke
1986; Edelson \& Krolik 1988; Robinson \& Perez 1990; and especially
P\'{e}rez et al. 1992).

To avoid the uncertainty in the interpretation of the peak of the CCF,
the centroid (center of gravity) of the CCF is often quoted.
In the infinite duration limit, the centroid of the CCF corresponds to
the centroid of $\Psi$ (e.g. Penston 1991; Koratkar \& Gaskell
1991a\footnote{The end result is correct, but the derivation contains
an error.}).
As Penston points out, this is intuitively obvious because the ACF is
an even function and the CCF is the convolution of $\Psi$ and the ACF.
The lag determined from the centroid of the CCF is therefore
sometimes preferred over the peak lag (e.g. Peterson et al. 1998a)
because it is more easily interpreted as the responsivity--weighted
radius of the BLR. Thus it has become common practice to quote both the
peak and centroid of the CCF when making lag estimations.

However, the centroid suffers from three serious flaws. 
First, the centroid of the CCF based on finite--duration light curves
is {\em not} identically the centroid of $\Psi$. This follows directly
from the fact that the sample CCF is not identically the sample ACF
convolved with $\Psi$, i.e., eq. (5) is only true in the infinite
duration case. The best that can be hoped for is that the centroid of the
CCF is approximately the same as the centroid of $\Psi$ if the
durations of the light curves are much longer then the lag.
Second, the centroid is poorly defined.
Obviously not all points in the CCF should be used to define the
centroid, only those near the peak should be included. So some threshold
is arbitrarily set, and the value of the centroid can depend very strongly
on what threshold is chosen (see Koratkar \& Gaskell 1991a,b). 
Third, because the centroid integrates over a range of values of
the CCF, it is more sensitive to the bias inherent in the CCF than
the peak. Thus the centroid underestimates the lag more than the peak.
This bias in the centroid then negates the argument that the centroid
gives a more characteristic BLR size than the peak. 
The problems with using the centroid to estimate the lag will be
illustrated via the simulations in \S 4, where it will be seen that
the centroid of the CCF generally fares worse than the peak
(see also P\'{e}rez et al. 1992).
\footnote{
A similar preference for the use of the CCF peak has been previously
shown by Wade \& Horne (1998). In using spectroscopy to measure radial
velocities, they found that fitting a very narrow Gaussian to the peak of
the CCF gives a more reliable velocity estimate 
than fitting a broad Gaussian to the CCF.}

\subsubsection {the removal of the mean }
The terms $\bar{x}$ and $\bar{y}$ in eq. (1) are the mean values of the
entire light curves. For a stationary process, the sample mean of all
the data is clearly the best estimate for the mean. But for a finite
duration realization of a stochastic process, this may not be the case. 
Scargle (1989) makes the point that for positive definite quantities (e.g.
fluxes), removing the sample mean is not always desirable. Instead,
removing a fraction of the mean may provide a better CCF estimate. The
question of what fraction to use depends upon the data themselves, and
experimentation (and judgment) is required to find what fraction is
optimal. One could in fact solve for the mean as a free parameter
(e.g. Press, Rybicki \& Hewitt 1992).

The standard definition of the CCF (eq. (1)) implicitly assumes the
time series are stationary in their mean and variance. To help fulfill
this requirement, one can ``pre--whiten'' the data by:
(a) removing a linear (or higher order) fit, or
(b) apply a differencing operator to the data. In the latter case, the
data to be analyzed are derived from differences of successive pairs of the
light curve: $x'_{i} \equiv x_{i} - x_{i-1}$. The differencing operator is
in fact a high--pass filter, and under high signal--to--noise
ratio (S/N) conditions it is the preferred technique to remove trends
--- see e.g. Koen (1993, 1994).
Unfortunately, it fails in practice because the S/N of currently available
UV/optical AGN variability data is too low, and one is left mostly with
noise.

\subsubsection{ the ``local'' CCF}
An alternative definition of the CCF in which only those $(N-k)$
points that overlap at a given lag ($\tau_{k}=k\Delta t$) are used to
determine the mean and standard deviations is:
\begin{eqnarray}
local \ CCF(\tau_{k}) \equiv
\frac{
\frac{1}{N-k} \displaystyle{ \sum_{i=1}^{N-k} }
(x_{i}-\overline{x_{*}})(y_{i+k}-\overline{y_{*}}) 
}
{ 
\left[\frac{1}{N-k} \displaystyle{\sum_{i=1}^{N-k}}
(x_{i}-\overline{x_{*}})^{2}\right]^{1/2}
\left[\frac{1}{N-k} \displaystyle{\sum_{i=k+1}^{N}}
(y_{i}-\overline{y_{*}})^{2}\right]^{1/2}
}  
\end{eqnarray}
where $\overline{x_{*}} = \frac{1}{N-k} \sum_{i=1}^{N-k} \ x_{i}$ \ and
\ $\overline{y_{*}} = \frac{1}{N-k} \sum_{i=k+1}^{N} \ y_{i}$ \ are the
means of $x_{i}$ and $y_{i}$ in the overlap interval.
We refer to this method of computing the CCF as the ``local CCF'' method.
The local CCF naturally accounts for the $1/N$ versus $1/(N-k)$
problem because only those values of the light curves which overlap at a
lag $k\Delta t$ are included in the determination of the mean and standard
deviations. The value of the CCF determined in this fashion is then
identical to the product--moment correlation coefficient, also
known as Pearson's $r$ statistic (see e.g. Press et al. 1996 for 
a discussion) and this is the method used by White \& Peterson (1994).
The simulations in \S4 will demonstrate that while the bias is not
completely removed, this definition produces a CCF with far more
desirable qualities. 
Notice that in effect the local CCF method applies a varying 
high--pass filter to the data whose cutoff frequency depends on the 
number of points in the time series at a particular lag (i.e. $N-k$).
Thus the local CCF handles data containing low frequency trends
far better than the standard CCF. 
For these reasons we prefer the local CCF over the standard
definition (eq. (1)), but we caution that CCFs constructed
in this manner are for lag determinations only, as to our knowledge
the statistical properties and the relationship between the local CCF and
the (a) inverse Fourier transform of the power spectrum or (b) 
coefficients of autoregressive (or moving average) processes has not been
investigated. Jenkins \& Watts (1969) do not recommend the local CCF for
these reasons; however, recovery of ARMA coefficients and the power
spectrum is not the goal of CCF analysis in the context of AGN echo
studies --- the determination of an unbiased lag estimate is.

\subsubsection { the CCF: an intuitive statistic? }
As Jenkins \& Watts (1969) point out, it should be kept in mind that
the standard CCF as defined by eq. (1) has not in any way been proven to
be the best possible estimator. It is used primarily because it is an
efficient, consistent estimator with intuitive appeal. However, our
intuition can be grossly incorrect in cases when $k$ is not $\ll N$.
So we should not consider equation (1) to be sacrosanct, and in
particular, the ``local ccf'' implementation  largely reduces the bias
problems mentioned above, and in our simulations, yielded results more
akin to our expectations than the standard  method.

The local CCF has an easy to understand interpretation (see e.g.
Bevington \& Robinson 1992): 
For a given lag $\tau_{k}$, there are $N-k$ overlapping points. 
A least--squares straight--line fit to the mean--subtracted values
of $y$ versus $x$ will yield some slope $b$. 
A non--zero slope suggests a correlation. However, the value of
$b$ cannot be used as a measure of the strength of the correlation
because $x$ and $y$ could be strongly correlated and yet have a
small slope (e.g., if $x$ spanned many orders of magnitude more than $y$
the slope would be small even if there were a perfect correlation).
Reversing the dependent and independent variables and fitting a line to
$x$ versus $y$ will give a slope $b'$. As with $b$, a non--zero slope $b'$
implies a possible correlation. The product $bb'$ is then a measure of the
correlation strength, independent of the slope of the relationship.
The quantity $\sqrt{bb'} \equiv r$ = local CCF($\tau_{k}$).

Despite the apparent intuitive appeal of using the CCF to detect a
correlation, a few caveats and comments are in order:
(i) equation (1) defines a {\em linear} correlation coefficient
between two series. The restriction to a linear correlation is arbitrary,
and a non--linear analysis may prove fruitful;
(ii) the use of non--parametric correlation tests (such as 
Spearman's rank--order correlation) may be of value
(see Press et al. 1996) since they tend to be more robust;
(iii) our intuition is based on the abstract mathematical case of
infinite duration time series, and this can be grossly misleading.

\subsubsection{ finite duration sampling}
Flagrant violations of our intuition about the ACF/CCF
can occur if the duration of the light curves are finite. The problem is
particularly serious in the AGN context because the lags of interest are
often a sizeable fraction of the total duration of the light curves.

In theory, the CCF should be the convolution of the ACF and the transfer
function $\Psi$. But even in the absence of noise and with ideal
sampling this will often not even be approximately true. Leaving aside
the effects of bias in the standard CCF definition, the major cause
of the difference between the expected and measured CCF is due to
changes in the {\em observed} ACF.
For any finite realizations of a stochastic process, the ACFs will
not be exactly the same, even if 
the underlying generating process is unchanged.
The usual interpretation of the CCF demands stationarity in the mean
and variance, but this is in general unlikely for finite duration
observations generated from a stochastic process, and in particular, AGN
light curves have red power spectra and are far from being stationary on
year--to--year timescales. As a consequence, the ACFs can vary from epoch
to epoch, and with it the CCF lags, despite no real physical changes in
the emission producing mechanism (either the continuum source or BLR). 
To help mitigate this effect, low frequency power in the data 
should be removed, or if the data S/N allow, a differencing operator
should be applied. In \S4.2 simulations will show the improvements in
the CCF lag determination that can result by removing low--frequency
trends and softening the edge effects in the finite light curves.

\subsection{ The errors in the lag determination }

Real data have finite time resolution,  contain noise, may be unequally
sampled, and have finite duration. The effects of finite sampling rate
are not a problem provided the data are not undersampled, and the effects
of noise and unequal sampling have been discussed extensively in the
astronomical literature, e.g., White \& Peterson (1994), Maoz \& Netzer
(1989), Peterson et al. (1998a), and references therein. Simulations
have shown that the CCF can be satisfactorally recovered under a wide
range of realistic sampling and noise conditions.
However, the effects of bias and finite duration observations have not be
properly appreciated when considering the errors in the lag determination.

Ignoring bias for the moment, there are two distinct sources of error: 
(i) external errors due to  observational noise and irregular sampling,
and (ii) internal errors due to the random nature of the light curves.
Finite duration sampling of the light curves brings about the latter
source of error. It is independent of observational noise or the sampling
pattern. 

Maoz \& Netzer (1989) estimated the errors on the CCF by
producing a set of simulated line and continuum light curves with
random sampling and noise, then constructed a ``cross--correlation peak
distribution'' (CCPD) histogram showing the spread in lags of the peak.
From the CCPD one can measure the mean (or median or mode) and the
associated uncertainty on the lag for the chosen model.
For their continuum time series they used either an interpolated version
of the Peterson et. al. (1985) Akn 120 light curve or the Clavel et al.
(1987) NGC 4151 light curve. So in fact {\em identical} parent continuum
light curves were used for each of their two sets of simulations. The
CCPD clearly shows a large spread in the determinations of the lag, but
this spread only shows the effects of the sampling and observational
noise. 

The White \& Peterson (1994) analysis is more general in that the
continuum light curves are not identical; instead they have a power--law 
power density spectrum (PDS) of either $P(f) \propto f^{-2.5}$ or
$\propto f^{-1.8}$, motivated by the power spectra of NGC 5548 (Clavel
et al. 1991) and NGC 3783 (Reichert et al. 1994), respectively. However,
although the continuum light curves are different in each realization,
they all have very similar ACFs (since by definition they have identical
PDS and the Weiner--Khinchin theorem states that the PDS and ACF
are Fourier pairs).
The scatter in their CCPD is therefore dominated by observational noise
and irregular and sparse sampling, and do not realistically include
finite sampling--induced changes in the ACF.\footnote {Technical note:
The construction scheme used by White \&
Peterson (1994) to add backgrounds to the
light curves (such that the fractional rms ``$F_{var}$'' matches the
observed value of 0.32 for NGC 5548) results in a correlation between the
intrinsic rms of the light curve and the amount of observational noise
added. The simulated noise is therefore not constant, nor is it dependent
only on the simulated fluxes; it also depends on the amount of intrinsic
variability. This correlation may result in a slight overestimate of the
reliability of the lag determinations, since for continuum light curves
with small intrinsic fluctuations the observational noise is reduced.}

In their pioneering work, Gaskell \& Peterson (1987) do indeed take into
consideration the changes in the ACF, since their simulated continuum
light curves were generated using a first--order autoregressive model
(e.g. see Jenkins \& Watts 1969). In fact their figure 12 shows 
the problems mentioned in the previous section: bias in the
correlation  coefficient (height of the peak of the CCF) and bias in the
position of the peak of the CCF (towards too small lags). However, the
emphasis of their work was to highlight their interpolation method for
coping with irregular, sparse sampling and observational noise, and
neglected the issues we investigate here.

The more recent analysis of CCF uncertainties by Peterson et al. (1998a)
attempts to generate a realistic CCPD using a combined Monte Carlo and
bootstrap method (see e.g. Press et al. 1996 for a discussion of the
bootstrap method).
The Monte Carlo ``flux randomization'' jitters the
observed data values by a random amount consistent with the
observational noise, while the bootstrap ``random subset selection''
picks subsets of the time series at random. The authors conclude that
the method produces estimates of the errors that are generous, i.e., 
slightly larger than the errors in the parent distribution. 
While their detection of a wavelength--dependent lag in NGC 7469 seems
irrefutable, the analysis of the errors is only partially complete.
In their simulations, the method used to generate the light curves 
was nearly identical to that in White \& Peterson (1994), so they do not
precisely mimic reality in the construction of the parent CCPD
distribution. The fact that the CCPD generated via a Monte Carlo and
bootstrap treatment of a single light curve realization is larger than
the parent CCPD is comforting, but this 
may still underestimate the true uncertainty in the lag.

Note that the Peterson et al. (1998a) implementation of the bootstrap 
omits roughly 37\% of the observed continuum and line data pairs; this
can be avoided if one reverses the order of the Monte Carlo flux and the
bootstrap sampling. By bootstrap sampling first, one can keep track of
the data pairs that are selected more than once and the error bars on
those points can be reduced accordingly, before the data are jittered by
the Monte Carlo method. This brings the method more in line with the
standard bootstrap technique (Efron 1983; Diaconis \& Efron 1983), where
in essence the data are not omitted, but rather, a random weighting is
given to the points --- in the analysis, a weight of zero is assigned to a
datum if it is not chosen, and a weight of $n$ is assigned if that datum
is selected $n$ times. The process is repeated many times and in all cases
the total number of input values used is constant.
Note that for data whose error bars are roughly constant, it is the
bootstrapping, not the Monte Carlo step, that gives the correct error
distribution, and for this reason the modification of the Peterson et al.
(1998a) technique is important. For the equal sampling case with equal
size error bars, Monte Carlo flux randomization alone can grossly
underestimate the CCPD width.

The Monte Carlo plus bootstrap method as suggested by Peterson et al.
(1998a) and modified to improve efficiency as stated above appears to be
the best way to 
estimate the CCPD and hence the uncertainty of the lag for a given time
series. Yet if the continuum light curve is not at least weakly 
stationary, knowing the uncertainty for a given realization does not give
a reliable estimate for the full range of scatter in the determination of
the lag. Even if the continuum generating process is stationary on
long timescales, short observations may mimic non--stationarity.
The fact that the observed yearly mean fluxes from NGC 5548 are not
consistent with each other is evidence that the process is not stationary
over the timescales of interest and so comparison of CCF from
different years is problematic. 
Only if the ACFs from year to year are identical can the changes in the
lag be ascribed to changes in the transfer function and hence changes in
the BLR.
In summary, even if large and apparently significant changes in the 
lag are observed, these do not necessarily imply changes in the
AGN/BLR --- the changes may simply be due to finite--length observations.

Returning to the issue of bias, the standard CCF will underestimate the
lag on average, and this bias will not be included in the uncertainty
estimates --- it is a systematic error.
Even with sufficient sampling and no noise, the CCPD is
skewed toward smaller lags, and this bias becomes worse as the
duration of the light curves decreases. The CCPDs shown by Maoz \& Netzer
(1989), and by Netzer \& Peterson (1997) do not show this skew because
they used a specific continuum ACF shape then resample it; there is a
bias in their CCPD, but it is nearly the same for each of their
simulations.  The local CCF method reduces the bias, but it is
still present.

There are a number of examples in the literature that show the presence
of the bias, e.g., the CCPDs in Litchfield et al. (1995) clearly show a
bias toward underestimating the lag in both the (local) interpolated CCF
method and in the discrete correlation function. Litchfield et al. (1995)
attributed this bias to the asymmetric shape of their simulated blazar
single--flare light curves (rise time much shorter than decay time), but
it is in fact a general property of the CCF.
The bias can also be see in Table 1 (column 6) of White \& Peterson
(1994) in which the results of Monte Carlo simulations show that the peak
of the CCF usually occurs at slightly smaller lags than the true 
lag.\footnote{Technical note: For the cases in White \& Peterson's Table 1
with anisotropic BLR cloud emission, i.e., the anisotropy factor $A=1$,
the values quoted for the true expected lag refer to the centroid of the
CCF. However, for the simulations it was the peak of the CCF that was
measured. For these asymmetric right triangle--shaped transfer functions,
the centroid and peak values are very different. So the comparisons for
cases with $A=1$ are only approximately valid, and the bias cannot be
readily noticed.}  
Another example of the presence of the bias can be seen in
the Monte Carlo tests listed in Table 2 (column 2) of Peterson et al.
(1998a) and the corresponding skewed CCPD shown in their Fig. 1.
This figure also illustrates that while the CCF centroid distribution 
can be more strongly peaked than the CCF peak distribution, it is also
more heavily skewed --- it has a higher precision, but lower accuracy.
Unfortunately, it is difficult to assess the amount of bias present in
real data since it depends on the true shape of the CCF, i.e., we need to
know the true ACF and the true $\Psi$ --- the broader either of these, the
worse the bias. Simulations are required to estimate the bias present in
the lag determination, and this then becomes model dependent.

\section{Simulations}
There is no doubt that the lag of the centroid (or peak) of the CCF for
the NGC 5548 H$\beta$ light curve is changing from year to year. The
question is, do these changes indicate that the transfer function $\Psi$
is changing, or do the changes merely reflect the changing ACF due to
finite duration sampling? To answer this question and to illustrate the
points made in \S3, we performed the
following simulations.

\subsection{Construction of the simulations}
In this investigation we consider only equally sampled data; this way
we cleanly separate the effects due to unequal sampling and those due to 
changes in the ACF. 
We construct the simulations such that the changes in the observed ACFs
are due to the finite length of the light curve, not due to intrinsic
changes in the AGN continuum--generating process, although this would
have the same effect.

To simulate the observed continuum, we created a time series from 
a simple power--law power density spectrum (PDS with $P(f) \propto
f^{\alpha}$)  with index $\alpha$=--2  and random phases.  A value of
$\alpha \sim$ --2 to --3 for the UV continuum in NGC 5548 has been
determined by Krolik et al. (1991), so $\alpha$=--2 is a reasonable
choice, though we emphasize that this is a convenient parameterization
for pedagogical purposes, not to be over interpreted as a true
representation of the AGN light curve. This artificial time series has
zero mean and spans 10 years in time with 1--day sampling.

To simulate the observed continuum light curve $C$, the time series
is normalized to give an intrinsic rms variability of 2.0 
(in units of $10^{-15}$ erg s$^{-1}$ cm$^{-2}$ \AA$^{-1}$), and
a constant value is added so the average continuum level is 10.0.
Gaussian distributed white noise with a standard deviation of 0.333 
is then added to mimic observational noise.

A Gaussian centered at $\tau$=20 d with width $\sigma$=6 d was chosen
for the transfer function $\Psi$.
This form of the transfer function is motivated by the appearance of the
observed H$\beta$ transfer function in  NGC~5548 (Horne, Welsh \&
Peterson 1991; Peterson et al. 1994), though we stress that the
conclusions derived from this choice of $\Psi$ are true in general. In
fact, because this $\Psi$ has a well--defined peak, unlike, say a
thick spherical BLR transfer function, our simulations present a somewhat
optimistic case because the resultant CCF should have a relatively sharp
peak. 

The line light curve $L$ is generated by convolving the original 
noise--free and zero--mean $C$ with $\Psi$. The line light
curve is then scaled to give an rms value of 1.5 (in units of $10^{-13}$
erg s$^{-1}$ cm$^{-2}$) and a constant of 7.5 is added.
Gaussian distributed white noise with a standard deviation of 0.30
is then added to the line light curves to mimic observational noise. 
The parameters are summarized in Table 1, along with the actual observed
values for NGC 5548  from Peterson et al. (1999).
Figure 1 shows a typical 10--year long simulated continuum light curve,
along with its local ACF and PDS.\footnote{Technical note: all PDS were
computed using linearly detrended and Welch tapered light curves. The
two lowest frequency bins were not used in the fit to the PDS.}
The local CCF between the simulated continuum and line light curve for
the entire 10 yr period is also shown.

The light curve is then broken into 10 isolated segments, each
300 days long. Each of these segments corresponds to a season's worth
of simulated AGN data. Note that each continuum light curve is
independent --- the mean, the rms variability, the ACF, and the PDS are
not fixed. Figure 2 shows the PDS and CCFs for five seasons
extracted from the light curve in Fig. 1. Notice the large changes in
shape and lag of the CCFs in these examples, all of which were taken
from the same parent light curve. Also notice the differences between the
standard and local CCFs.

To build up a statistical number of realizations, the above construction
process was repeated 100 times, yielding 1000 seasons of independent
simulated continuum and line light curves. The standard and local methods
were used to compute the CCF for each of the 1000 pairs of light curves.
Both the peak of the CCF and the centroid position were recorded;
as in Peterson, et al. 1999, only values of the CCF that lie above 0.8
times the maximum r value were used in the centroid determination. 

\subsection{Simulation Results}
\subsubsection{The local vs. standard CCF}
Figures 3 and 4 show the peak lag values for each of the 1000 trials,
along with their CCPD, i.e., a histogram of the lag values.
Results from the local CCF method (Fig 3), and the standard CCF definition
(Fig 4) are shown. The superiority of the local method is immediately
evident. These figures also illustrate two points: (1) there is a
downward bias in the CCF, so that the average or median value
underestimates the true lag of 20 d; 
(2) the scatter of the CCF peaks is very large.

In Fig. 5, eight different CCPD are shown, each resulting from a
different method of computing the lag, but all from the same 1000
pairs of light curves. 
On the left, the peak and centroid lags are shown for the local and
standard CCF methods. The right panels show the same but for
light curves that have had a linear fit subtracted prior to computing
the CCF. This figure reveals several features: 
all CCPD show a downward bias, and this bias is very severe for the
standard CCF method; the centroid distributions are more susceptible to
bias than the peak distributions; the linear detrending greatly improves
the standard method, but only slightly improves the local method (because
the local method inherently contains a high--pass filter). 
The large number of CCFs that peak at zero delay in the standard CCF are
due to trends in those season's light curves: a strong linear trend in
both continuum and line light curves will dominate the CCF.  Linear
detrending of each season's light curves is therefore highly beneficial.

The simulations described above are optimistic in several regards:
(i) the light curves are equally sampled with no gaps, (ii) the
observational noise--induced scatter
in fluxes are purely independent and Gaussian, (iii) the transfer function
has a well--defined peak, and (iv) the duration of the light curves are 
15 times longer than the lag of the peak of the transfer function. As a
result, the conclusions drawn from these simulations are robust ---
more realistic simulations would show a larger scatter.

It was noticed that on occasion, the line ACF was narrower than the
continuum ACF. This has sometimes been seen in AGN light curves and
suggests that $\Psi$ contains negative values or is non--linear (see
the discussion by Sparke 1992). However, it can simply be a
side--effect of finite duration sampling.

\subsubsection{The effects of the continuum power spectrum ``color''}
To test the sensitivity to the assumed PDS power--law exponent $\alpha$,
we carried out simulations using parent continuum light curves with a
$1/f$ and $1/f^{3}$ PDS. The peaks of the local CCFs were determined,
after the light curves had linear trends removed. The results are shown
in Figs. 6 and 7, where it
is clear that the scatter in the lags depends strongly on the value of
$\alpha$. This is because the width of the ACF is sensitive to 
$\alpha$: redder PDS (more negative $\alpha$) have broader ACFs.
As stated in eq. (5), the ideal CCF is identical to the
transfer function $\Psi$ convolved with the ACF, so a broad ACF yields a
broad CCF whose peak is poorly defined. The consequence is that the redder
the continuum fluctuations, the more the ability to infer properties of
$\Psi$ from the measured CCF degrades. 
These simulation can be compared with the three model CCFs shown in
Fig. 4 of Edelson \& Krolik (1988), where redder continuum light curves
yield a stronger correlation, but contain less structure and hence
less information.

The above claim that the CCF peak should be more easily measured for
whiter PDS leads to an apparent contradiction. The transfer function acts
similarly to a low--pass filter, hence high frequency variations present
in the continuum are averaged out and are not seen in line light curve.
The thicker the BLR, the more the high frequencies are lost. This
suggests that a continuum light curve dominated by low frequencies, i.e.,
a very red PDS, would provide a better ``driver'' for echo mapping.
Indeed, this effect is seen by White \& Peterson (1994) in their
CCPD comparisons using $1/f^{1.8}$ and $1/f^{2.5}$ simulated continuum
light curves. The contradiction is resolved by noting that the  
signal--to--noise ratio of typical AGN variability data is rather
low, hence noise in the light curves is important. For continuum light
curves with equal intrinsic rms variability but different PDS power--law
slopes, the deleterious effect of white noise is more pronounced for
whiter PDS. In other words, the S/N is timescale dependent: low
frequency variations have effectively a higher S/N than high frequency
variations. Since convolution with the transfer function preserves
low--frequency power, continuum light curves with redder PDS yield CCFs
that are are less sensitive to noise. However, one cannot escape the fact
that a very red PDS continuum will have a very broad ACF and CCF, making
its peak and centroid determinations difficult. For high S/N data,
a whiter PDS will enable a richer echo mapping analysis.

\subsubsection{The duration of the light curves}
As the duration of the time series increases, one expects the
reliability of CCF lag determination to improve. To quantify this
behavior, Figure 8 shows the mean and median lag values plotted as a
function of the length of the hypothetical observing campaign. 
Five curves are drawn, corresponding to the mean and median for the
peaks and centroids of the local CCFs, and the median of the
peaks using the standard CCF method. As before, 1000 simulations were
used to determine the mean and median, and these simulations were
identical in all respects except for the number of points.
While all the distributions encompass the true value within $\pm
1\sigma$, they are all biased too low.  For simulations that match the
characteristics of the observations of NGC 5548, the lag is
underestimated by $\sim$5\% using the local CCF method.

From this figure it is clear that the median is a far better statistic
than the mean. This is because large outliers are not rare. 
For light curves of 150--300 days duration, the median bias in
the local CCFs is roughly 5--10\%. For shorter duration light curves,
the bias and variance increase rapidly: for 100 day long time light
curves the bias is $\sim$10\% and for 50 day long light curves the
bias grows to $\sim$25\%. For light curves this short the huge
uncertainty in the lag makes a single estimate almost meaningless. 
For comparison, the standard CCF method produces significantly more
biased values: even with 300--day long light curves the bias is
$\sim$25\%. 

The commonly held belief that the time series used to compute the CCF
should be {\em at least} 4 times longer than the lags of interest, and
preferably $\sim$10 times longer, is illustrated in this figure.
As the light curves lengthen, the lags only asymptotically approach
the unbiased value. Extending a 200 day long light curve by 100 days
does not decrease the bias nearly as much as extending a 100 day long
light curve by the same amount. Once the light curves exceed about 10
times the lag, increasing the S/N of the data leads to
more significant improvements then extending the duration of the light
curves.

\subsubsection{The signal--to--noise ratio}
The previous simulations were all based on a line S/N of 5, mimicking
the NGC 5548 H$\beta$ observations. S/N is defined here as the ratio of
{\em intrinsic} line rms variations to the simulated observational
noise per datum --- see Table 1 for the numerical values.
To quantify the effects of a change in S/N, Fig. 9 shows the median
of the CCPD for S/N values of 2.5, 5 and 10, plotted as functions of
the duration of the light curves. For this figure, lag is defined as
the peak of the local CCF. Both the line S/N and the continuum S/N
were boosted by the same factors, achieved in realization by reducing
the amount of added simulated observational noise.

As expected, the higher the S/N, the smaller the bias, and more
importantly, the smaller the variance about the median. From this
figure it can be deduced that under certain conditions, doubling the
S/N of the individual observations can be as significant as doubling
the duration of the light curves. 

\subsubsection{The effects of detrending}
In \S3.1.4, 3.1.7, and 4.2.1 it was stated that ``detrending'' or
removing low frequency power from the light curves can improve CCF lag
determinations. Figure 10 explicitly shows this effect. Plotted are the
median values of the 1000 CCF simulations versus the order of the
polynomial used to remove trends from the light curves (order 0 = mean,
order 1 = linear, order 2 = quadratic, etc.). The detrending was
accomplished by least--squares fitting a polynomial to the entire
light curve for the observing season (300 days in all cases) then
subtracting off the polynomial prior to computing the CCF.
In all cases identical light curves were used (with a line S/N = 5).
For clarity, only the error bars for the median lag computed with
the local CCF are shown.
In general, as the order of the polynomial increases, the bias in the
CCF decreases. 

For the standard method, simply removing a linear trend can result in
a substantially better estimate of the lag. Significant additional
improvements can be made by going to a 3rd or 4th order polynomial.
However, for higher orders, the variance increases, offsetting the
benefits of detrending. 

The beneficial effect of detrending is more pronounced for the standard
CCF method than for the local method because the local method
intrinsically contains a detrending--like filter (see \S 3.1.5).
Nevertheless, the local method also benefits from low--order
polynomial detrending of the light curves. (The apparent degrading of
the local CCF median when going from no detrending to linear
detrending is a statistical event.
More typically, the median value with no detrending is equal
to or worse than with linear detrending.)
Figure 10 again illustrates that the local CCF outperforms the
standard CCF, and that the peak is a better (less biased) estimator 
than the centroid. However, the local CCF is more noisy than the
standard CCF, particularly at large lags, and this grows worse with
higher order detrending.\footnote{In the situation where outliers
become common at the largest lags the median is no longer a good
statistic to use to characterize the CCPD, since it depends on the
limits of where the lags are computed (i.e., the median becomes
sensitive to the endpoints of the interval over which the CCF is
computed). A better statistic would be the mode or the center of a
narrow Gaussian fit to the distribution.}

Conceptually, detrending works for the following reason: AGN light
curves have a red power spectrum, so most of the signal in the light
curves are contained in the lowest frequencies. However, these low
frequencies are the most poorly sampled in the light curve: there is
only one measurement of the lowest Fourier frequency, two of the
second lowest frequency, and so on. Due to the small number of
samples, statistical fluctuations are very important. This is in
contrast to the high frequency power, where there are many samples,
but the observational noise is large (or even dominant). 
Detrending the light curves with polynomials applies a smooth and
gently rolling high--pass filter.\footnote{For equally sampled data a
sharp, well--defined high-pass filter could be applied in the Fourier
domain, but for unequally sampled data working in the Fourier domain
is problematic.}
The CCF will no longer be dominated by poorly sampled low frequencies,
and hence less prone to random fluctuations.  The scatter in the CCF
is therefore reduced. Another way to think of it is that the detrending
sharpens the ACFs, and since the CCF is approximately the ACF convolved
with the $\Psi$, the CCF sharpens up.

To understand why detrending reduces bias, one must realize that
the CCF is extremely efficient at finding the lag if the time series PDS
are white; otherwise the CCF can give poor results. For example, 
if the time series contains a trend then for a long time interval the
data will tend to be above (or below) the mean. Thus the data  
values are not randomly distributed about the mean; instead they are
highly correlated on long timescales. This correlation will dominate the
CCF and the peak of the CCF will occur at (or near) zero lag.
Thus there is a bias towards small lags if there is any low frequency
power in the time series.
For example, the peak of the standard CCF will occur at zero lag for
any two linear light curves. Unless the deviations from the straight
lines are large, the CCF will tend to peak at zero lag. 

AGN light curves are dominated by low frequency power hence the CCF
will be biased towards too small lags unless the data are
``pre--whitened''.\footnote{Determining radial velocity shifts of lines
in a flux spectrum does not suffer as much bias because the power 
spectrum of the flux spectrum is mostly white. Nevertheless, any
trends in the continuum must be removed or the resulting radial
velocities will be biased.}
Detrending the light curves via polynomials is one way of removing low
frequency power; other methods include subtracting off splines or a
moving average, applying a differencing operation (see e.g.
Chatfield 1996), or directly multiplying by a high--pass filter in the
Fourier domain. Since AGN light curves are red, it is clear that {\em
some} form of pre--whitening should be carried out. 

What order polynomial should be used to detrend the light curves?
The answer depends on the characteristics of the data:
the redder (more negative $\alpha$) the power spectra, the more
detrending is required; the lower the S/N, the less detrending can
be tolerated. 
AGN light curves have power at all observed frequencies so there will
always be linear and other low--frequency trends, independent of the
duration of the light curve. The minimum order of the detrending
polynomial is therefore insensitive to the length of the light curve:
a linear trend should always be removed. A crude estimate for the 
maximum order can be made as follows.
A polynomial of order $M$ has at most $M$ zeroes, so it removes
power on timescales greater than $\sim 2T/(M-1)$ where $T$
is the duration of the time series.
For a reliable CCF estimate, a light curve with a duration of 5--10
times the lag timescale is necessary. This gives a polynomial of order
$M\sim$ 0.4--0.5 $\times \ T/\tau_{\Psi}$, where $\tau_{\Psi}$ is
the lag expected for the given transfer function. If the S/N is poor, the
maximum order polynomial for detrending will be less than this. 
When the light curves are heavily detrended, much of the intrinsic
signal in the data is removed, leaving lower and lower S/N data for the
CCF to work with.  The lag of the peaks of the CCFs will therefore not
necessarily converge with increasing detrending. In the extreme limit
where the filtering leaves only the observational white noise, the ACF
again will peak at zero lag because of correlated noise between the
continuum and line observations (since they are both measured from the
same spectrum). So while the detrending removes bias, it also increases
the variance, and in extreme limits it re--introduces a bias. For this
reason, large amounts of detrending is not beneficial. The technique of
differencing the data (see \S 3.1.4) removes {\em all} low--frequency
trends, and hence is not a viable option for data that is not of
exceptionally high S/N. 

The strong recommendation that results from this work is that
removal of low--frequency trends in the light curves can significantly
improve the reliability of the CCF lag determinations. Removal of a
linear trend is essential; removal of a cubic or quartic trend is
recommended; higher orders may be useful if the correlation remains
strong enough to provide an unambiguous determination of the peak.
In practice, one should compute and compare the CCF for progressively
larger amounts of detrending. 

\subsubsection{The effects of tapering}
Tapering (also called ``windowing'') a time series is common practice
in Fourier analysis (see, e.g., Jenkins \& Watts 1968; Press et al.
1996). Tapering helps compensate for ``end effects'' of a finite
duration time series:
a sampled time series can be thought of as the product of two time
series, the ``true'' infinite duration time series and a time series
whose value is unity during the data acquisition interval and zero
elsewhere. The multiplication of the true time series with this sampling
function is identical to convolution of the Fourier transform of the
true time series with the Fourier transform of a boxcar.
The result is that the Fourier transform is broadened by convolution
with a sinc function. The broadening results in ``leakage'' of power
from one frequency into other frequency bins. 
Tapering the light curves consists of multiplying the light curves
with a function that slowly goes from zero to unity and back over
the duration of the experiment. The new time series has ``softer''
edges that produce a Fourier transform with less leakage.
For time series that have a red power spectra,
the leakage of power from low frequencies to higher frequencies
is significant, and possibly even dominant at if the intrinsic PDS is
redder than $1/f^{2}$.
Because of the equivalence of the CCF with its discrete Fourier
transform counterpart, reducing leakage from low to high frequencies
should improve the CCF lag estimate.
The red PDS of AGN light curves means leakage is significant and
suggests that tapering the light curves can have a benefical effect.

To explore the possible benefits of tapering, simulations were carried
out using light curves that were detrended and tapered prior to
computing the CCF. 
The taper was applied in two ways: (1) a global taper, applied once to
the entire light curve; (2) a local taper, applied to each overlapping
segment pair. The global taper is more akin to what is used to reduce
spectral leakage; the local taper forces the same taper to be applied
independent of lag.

The results indicate that tapering does on average improve the
reliability of the CCF lag estimate. However, the effect is small
compared to the effect produced by detrending. As expected, the greater
the amount of detrending, the less effect the tapering had. The two
different methods of tapering (global and local) produced similar
results for small lags; the differences were much less than the
variance in the estimates of the lag.
For globally tapered light curves, the local and standard methods of
computing the CCF gave very similar results for small lags. 
For large lags, the global taper significantly reduces the amplitude of
the CCF. This has two effects: (1) it greatly reduces the number of
outliers in the CCPD; (2) it introduced a strong bias against finding a
correlation at a large lag. Provided the light curves are long compared
to the true lag (something that is not known a priori), the latter
effect is not serious. In summary, tapering does have a beneficial
effect and the benefits are not very sensitive to the specific method
of tapering (or taper shape), but the effect of detrending is far more
important.  In practice, one should compute the CCF several ways: using
the standard and local method, different amounts of detrending, and
with and without tapering.

\section{Discussion and Conclusion}

We have discussed some properties of the cross correlation function,
specifically in the AGN echo mapping context. The two main issues we
address are (1) the bias in the CCF and (2) the uncertainties in the CCF
lag determinations. Both of these stem from finite duration sampling of
the light curves, not irregular/sparse sampling or observational noise.
Bias can also be introduced if low frequency power dominates the light
curves. Since AGN light curves have a red power spectrum, this second
source of bias is also present.

Because of the bias problem the CCF fails {\em on average} to reproduce
the correct lag.  The bias is inherent in the definition of the 
standard CCF itself and depends strongly on the ratio of the intrinsic
(true) lag to the duration of the observed light curves and also the
sharpness of the continuum autocorrelation function and transfer
function. Unfortunately the amount of bias cannot be determined from
the data themselves, i.e., one needs to know the true CCF in order to
calculate the bias. As a result, exact corrections are impossible and
simulations are required to estimate the statistical size of the bias.
However, much of the bias can be removed by simply detrending (and
to a lesser extent tapering) the light curves.

The impact on AGN variability studies is that the standard CCF tends to
underestimate the true time lag, therefore the derived characteristic
radius for the BLR is underestimated. From simulations designed to mimic
the well--sampled light curves of NGC~5548, the estimated lag is too low
by $\sim$ 5--10\%; for more poorly sampled light curves the bias can
be much larger. This bias amplitude is based on using the ``local'' CCF
method, in which the means and standard deviations used to calculate the
CCF are determined using only those parts of the light curves that
overlap at for a given lag. We find that the local CCF gives superior
results compared to the standard definition of the CCF, where the bias
can be 3 times larger. 
Although the size of the bias is relatively small compared to the
intrinsic uncertainty in the measured lags of many AGN, as the quality
of the data continues to improve, the bias will not be negligible and
its effect should not be ignored.

We also find that the lag of the centroid of the CCF does not 
yield a more accurate representation of the BLR size because
(1) it is more heavily biased than the peak of the CCF; (2) unlike the
infinite case, the centroid of the sample CCF does not necessarily
correspond to the centroid of the transfer function.

It has been observed that the H$\beta$ lag in NGC 5548 changes from year
to year (Peterson et al. 1999) and this can be interpreted in several
ways. The variations can be attributed to the AGN itself, e.g., the BLR
structure may be evolving, or the illumination of the BLR by the
photoionizing source may be changing (Wanders \& Peterson 1996), or the 
engine producing the continuum variability is changing such that the
continuum ACF is variable. However, an alternate explanation is 
simply that the changing CCF lag is due to finite duration sampling of
the light curves. Simulations that mimic the optical continuum and
H$\beta$ observations of NGC~5548 demonstrate that, even with perfect
sampling and with a transfer function that has a well--defined peak, the
scatter in measured CCF lags is large.
Thus the scatter in the observed $H\beta$ lags in NGC~5548 can
be attributed to finite duration sampling of a random process.

Observations have shown that AGN flux time series are not stationary on
timescales spanning several observing seasons, i.e., the means and
variances of the light curves do not remain constant from year to year.
Since the continuum variability properties are not constant (in
particular the ACFs), one cannot use the observed CCFs to unambiguously
deduce changes in the transfer function.

It is of course possible that the changing lag is intrinsic to the AGN,
but we have shown that the scatter in the lags are also consistent with
the interpretation of being a consequence of finite duration sampling of
a random walk--like process. Our simulations produce
a distribution of lags that is as wide as the observed scatter. Given that
the artificial data were oversampled, equally sampled, had perfectly
known noise characteristics and a transfer function with a well--defined
peak, the results of the simulations are highly robust.

To determine if the observed lag variations are intrinsic to the AGN,
one needs to show that a realistic simulation produces a narrower scatter
in lag distribution than what is observed, or that yearly changes
in the lags are not random. Given that much longer observing runs than
what has already been obtained for NGC 5548 are not feasible, the
resolution of the question of the significance of the changing lags will
demand new data with much higher S/N. This would
substantially tighten the scatter in the simulated lag distributions,
while its effect on the observed scatter depends on if the variations are
intrinsic or not. Also, a better understanding of the continuum
variability characteristics such as the power spectrum power--law exponent
$\alpha$ would allow more realistic simulations. As we have shown, the
scatter in the simulated lag distribution depends strongly on the power
law slope of the power spectrum. In this regard, a Fourier analysis of the
long--term NGC~5548 continuum light curve is warranted.

Finally, enumerated below are some practical suggestions that can
improve the reliability of CCF lag determinations in AGN:
(1) Detrending the light curves produces far more reliable CCFs. 
Linear detrending is required, and higher order detrending can be
beneficial if the S/N is high;
(2) The peak of the CCF gives a more reliable lag estimate than the
centroid.
(3) Tapering the light curves also has a beneficial effect, though not as
significant as detrending;
(4) The ``local CCF'' is less biased and therefore gives better results
than the standard CCF, especially for small lags. However, if the light
curves are detrended and tapered, the advantage the local CCF has over
the standard CCF is small.
If simulations are used to estimate uncertainties in the lag estimates,
then: (5) The median of the CCPD is more reliable than the mean or mode
for light curves that are not too heavily detrended; 
(6) An improvement of the bootstrap + Monte Carlo method (Peterson et
al. 1999a) as described in \S3.2 should be used. However, simulations of
this nature can only yield an estimate of the uncertainty of the lag for
that particular sample of light curve. Without an understanding of the
true ACF itself (not the sample ACF), estimates based on resampling or
perturbing the observed sample light curves can underestimate the
variability of the lag.

\acknowledgments
The author thanks an anonymous referee whose comments led to a
significant improvement in the depth and presentation of this work.
The author wishes to express appreciation to E.L. Robinson for
extremely valuable discussions and to Chris Koen for inspiring this
investigation and for comments on a draft of this paper.
The author also thanks Divas Sanwal for helpful comments throughout 
the various stages of this work.
The author acknowledges with gratitude the work of the many members of
the AGN Watch and the availability of the data and papers on their web
site: ``http://www.astronomy.ohio-state.edu/\~{}agnwatch/''.
This work was supported through NASA ADP grant NAG5-7002.


\clearpage

\begin{figure}
\epsscale{0.8}
\plotone{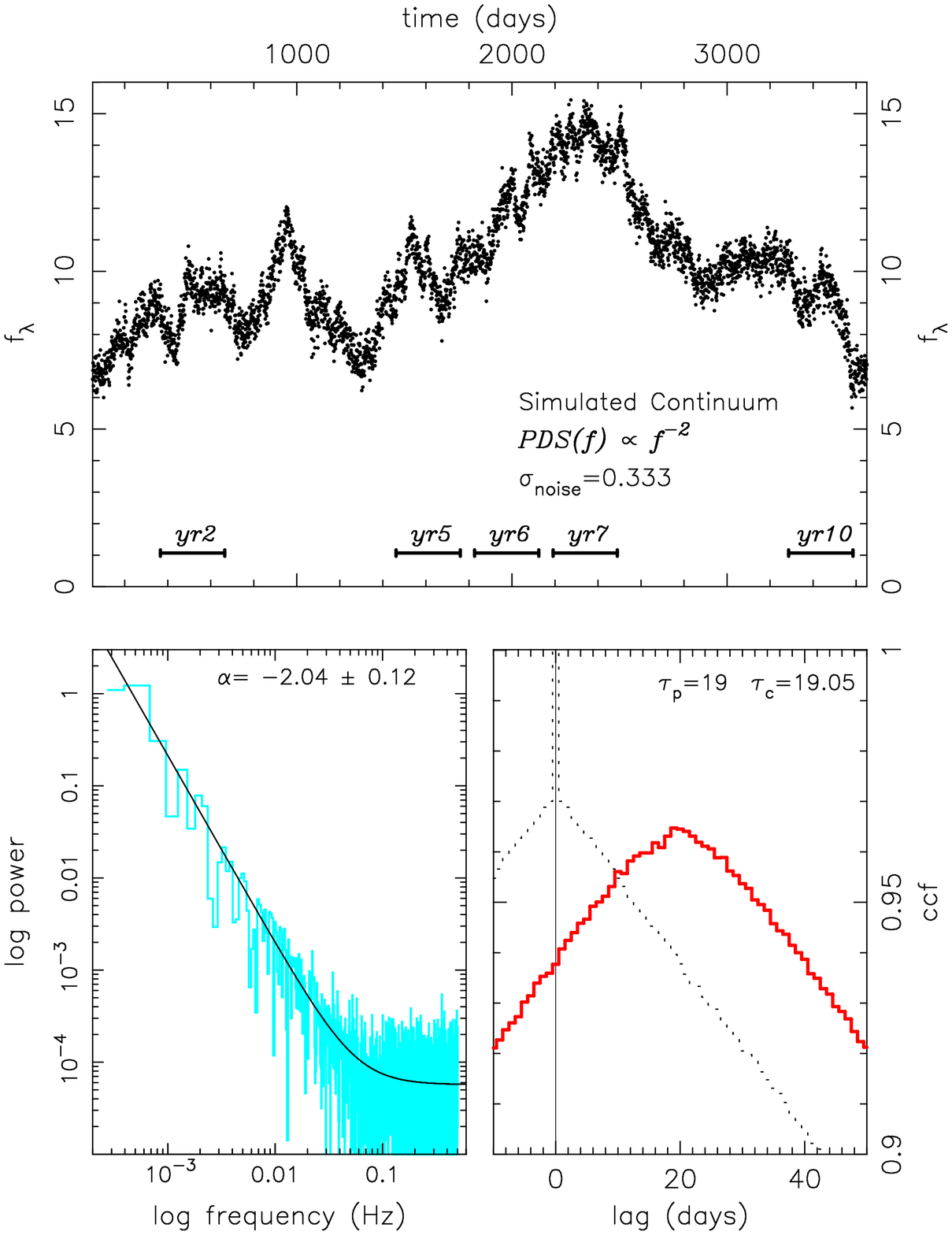}
\caption{ {\it upper panel:} Simulated optical continuum light curve
spanning 10 years. The simulation was designed to closely match the
observed characteristics of NGC~5548 in terms of variability amplitude,
observational noise and an intrinsic $f^{-2}$ power--law power spectrum.
Five different 300 day long observing seasons are marked.
{\it lower left:} The power density spectrum of the light curve shown in
the upper panel, with a power--law plus white noise fit.
{\it lower right:} The ``local'' cross correlation function of the
simulated observed continuum light curve shown in the upper panel and
line light curve (not shown). The transfer function used to generate the
line light curve is a Gaussian centered at 20 days and with a width of
$\sigma$=6 days. Also shown is the continuum autocorrelation function
(dotted line).
\label{fig1}}
\end{figure}

\begin{figure}
\epsscale{0.7}
\plotone{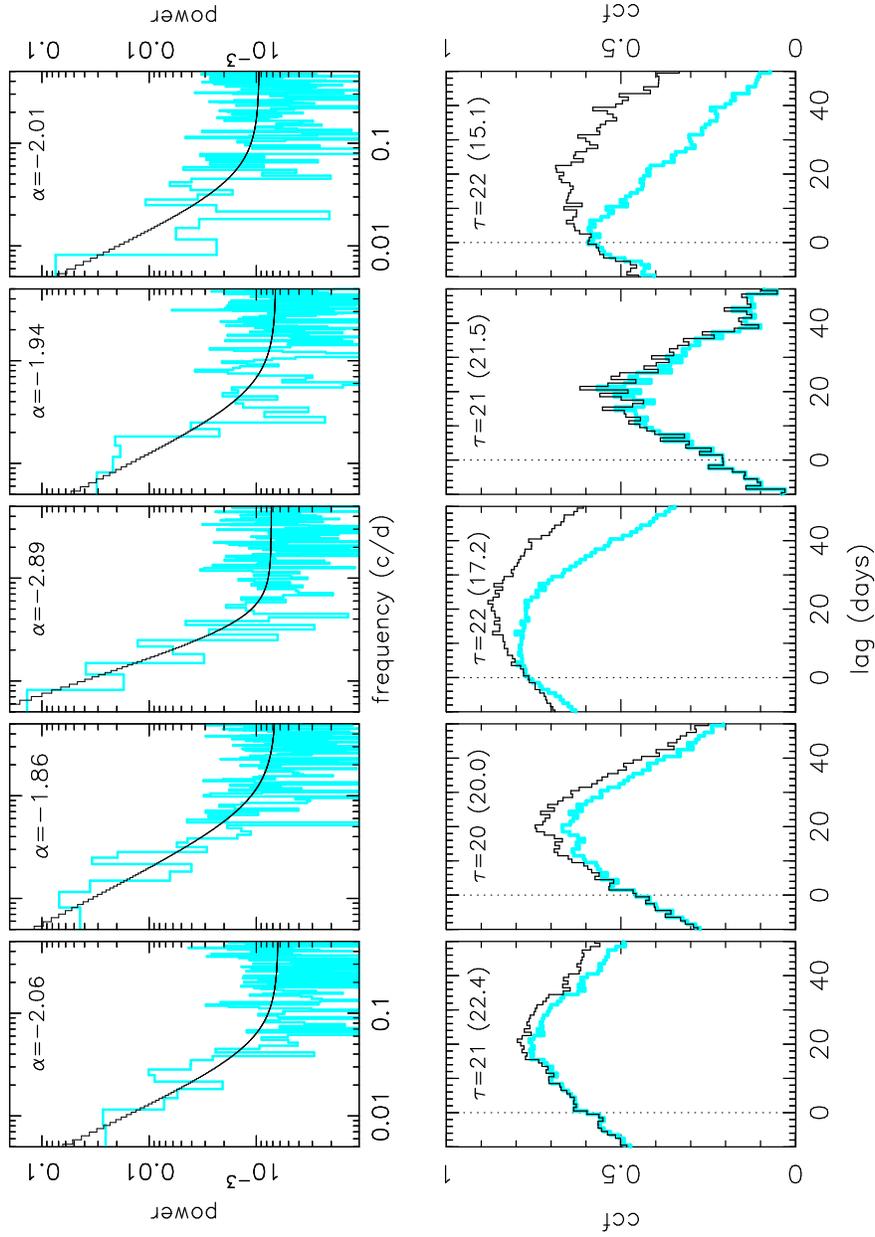}
\caption{Continuum power spectra and cross correlation functions
for the five seasons of simulated data marked in Fig.~1. The fit
and measured power--law index $\alpha$ for each of the power spectra are
given. Both the local CCF (dark) and standard CCF (light) are shown.
The lags as determined by the peak and the centroid of the local
CCF are quoted (the centroid value is in parenthesis).
Note the differences between the seasons, despite all being derived 
from the single continuum parent light curve shown in Fig 1.
Also note the rather extreme differences between the local and standard
CCFs.
\label{fig2}}
\end{figure}

\begin{figure}
\epsscale{0.7}
\plotone{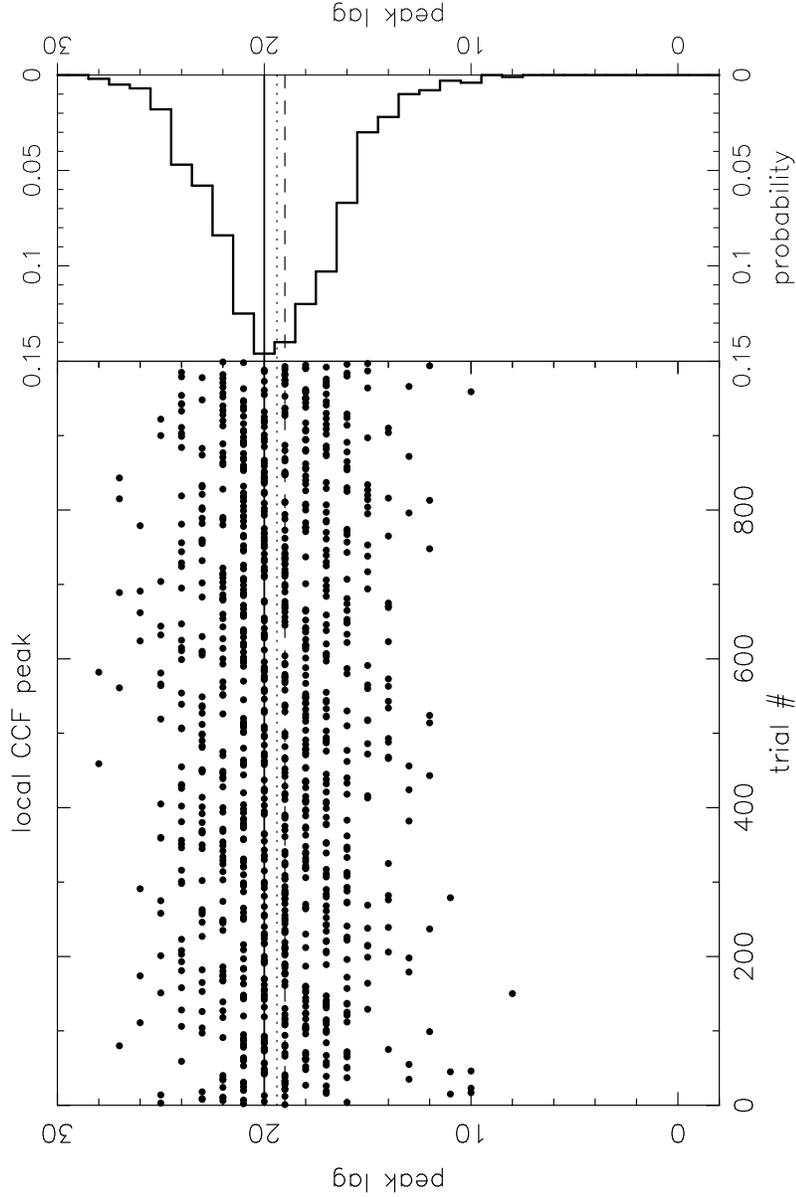}
\caption{The results of 1000 measurements of the CCF lag based on
simulations designed to match the optical continuum and H$\beta$ light
curves of NGC~5548. The left--hand panel shows the individual lag
determinations, using the peak of the ``local'' CCF to determine the
lag. The true simulated lag is 20 days; the mean and median of
the distribution is shown by the dotted and dashed lines.
The right--hand panel shows a histogram of the lag distribution. 
Note the bias --- the distribution is biased too low.
\label{fig3}}
\end{figure}

\begin{figure}
\epsscale{0.7}
\plotone{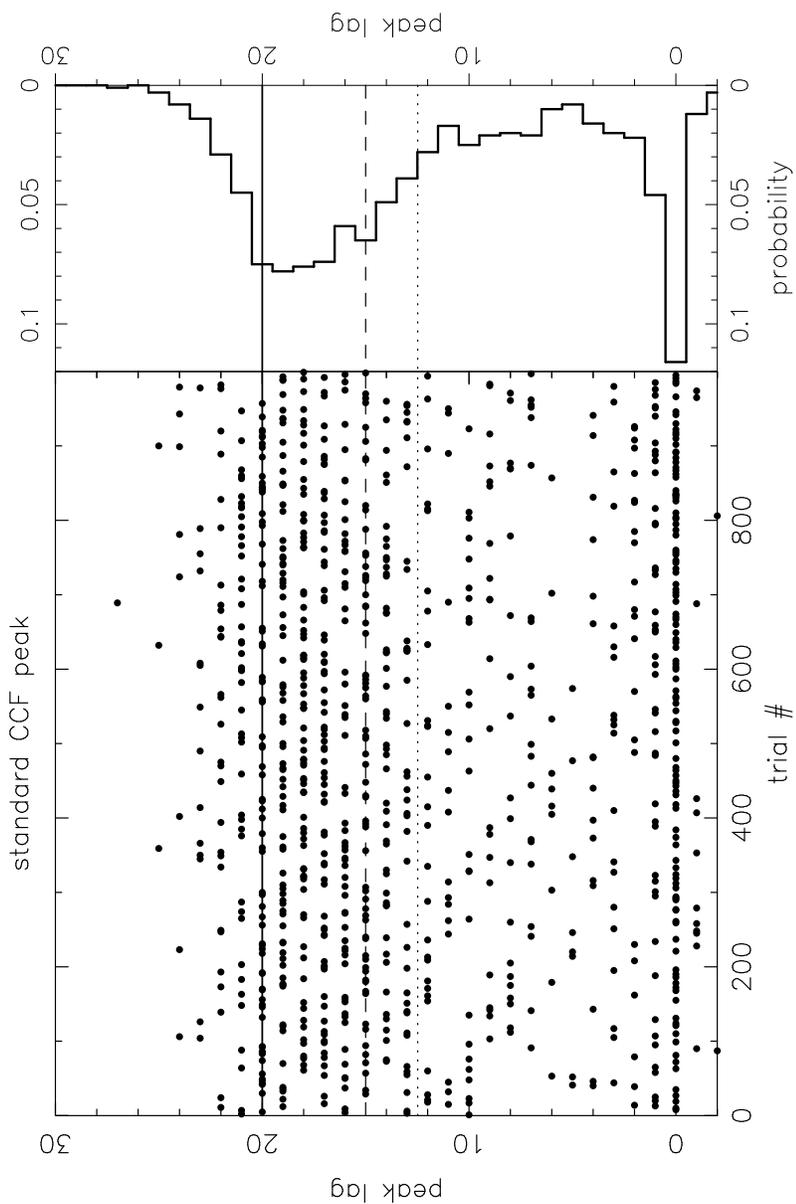}
\caption{Same as for Fig. 3, except the standard CCF was used. 
Notice the much larger scatter in the distribution, and the far worse
bias. Ths spike at zero lag is due to the presence of low frequency power
dominating the light curves.
\label{fig4}}
\end{figure}

\begin{figure}
\epsscale{0.7}
\plotone{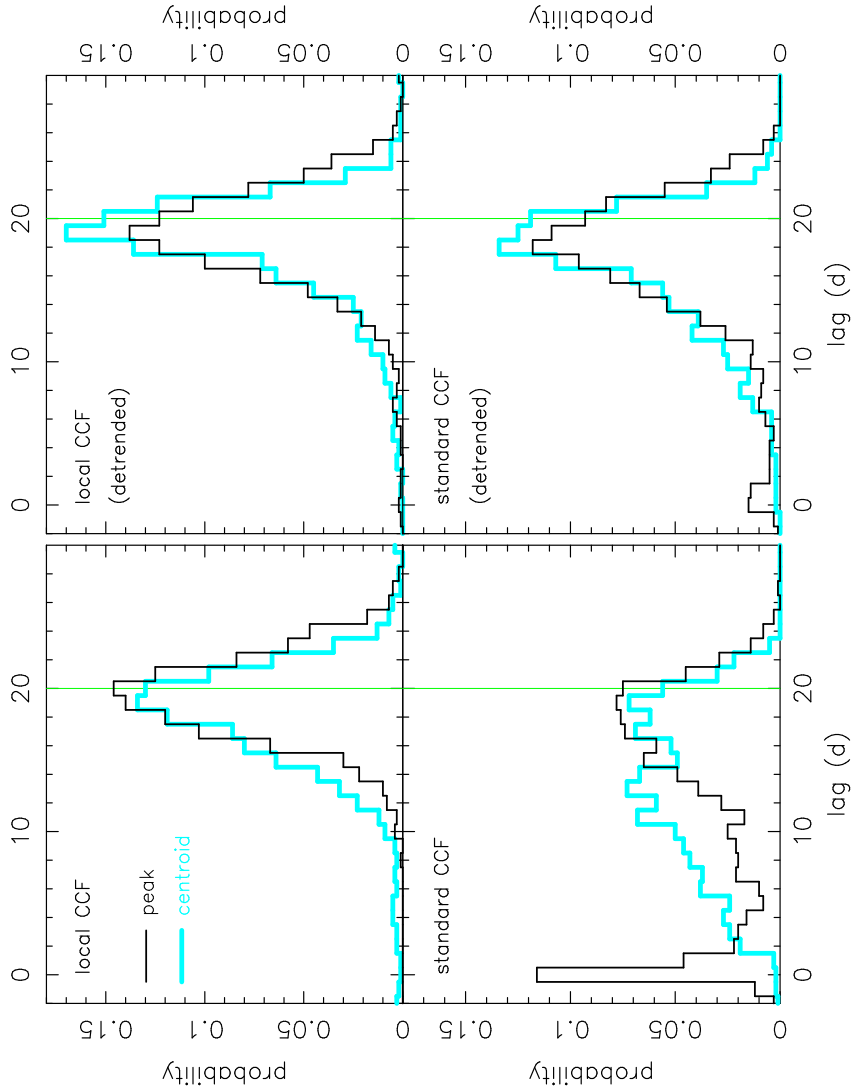}
\caption{The four panels show the CCF distributions based on different
methods. The darker line shows the distribution of the peaks of the CCF,
the lighter line is based on the centroids. The upper panels show
the CCF
computed using the ``local'' definition, while the lower panels are
based on the standard CCF. The left--hand panels use the light curves
with the mean removed prior to computing the CCFs; the right--hand panels
used light curves that had a linear fit removed prior to computing the
CCFs. In all cases, the exact same light curves were used. Notice the
superior quality of the local CCF method. Linearly detrending the
light curves improves the standard CCF significantly, but it is still not
as good as the local method. In all cases, a bias towards too small a
lag is present.
\label{fig5}}
\end{figure}

\begin{figure}
\epsscale{0.7}
\plotone{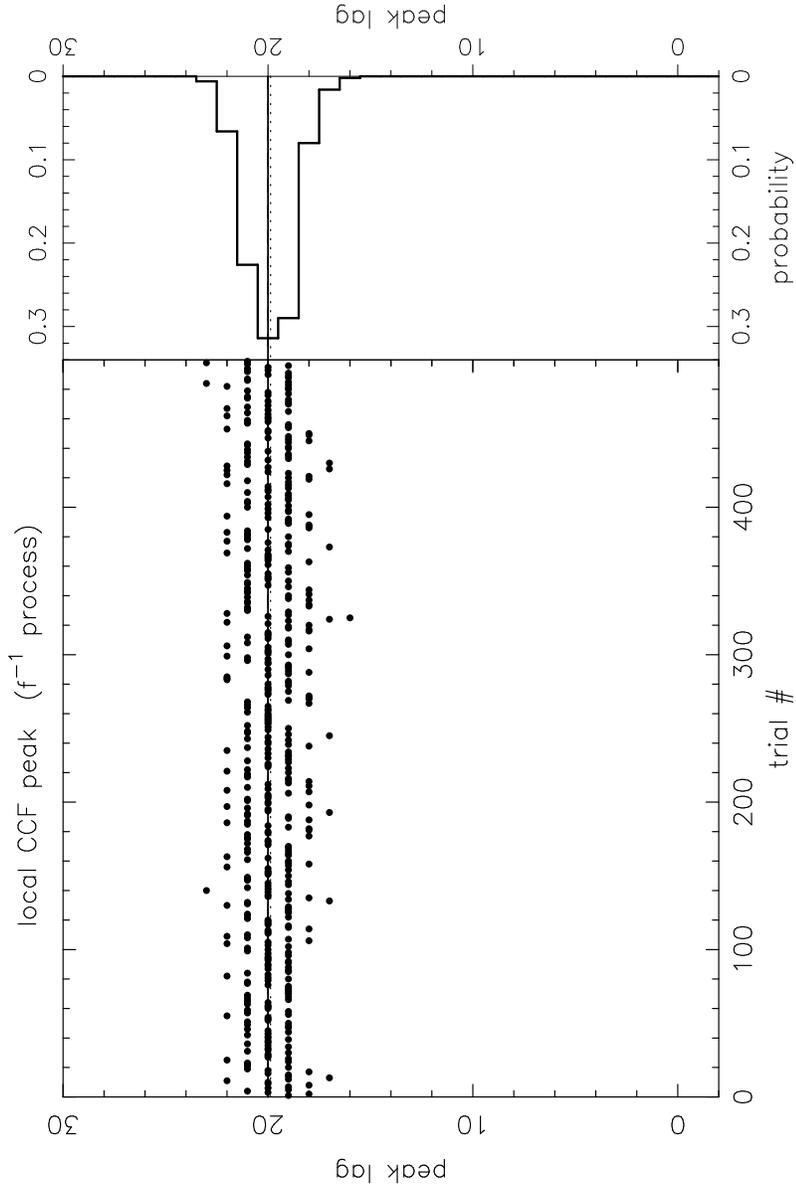}
\caption{The results of 500 measurements of the CCF lag based on
simulations using a $f^{-1}$ power--law power spectrum. The same noise
and sampling was used as with the simulations shown in Fig.~3. The smaller
scatter is due to more information content and narrower ACF in this 
``whiter'' continuum light curve, not due to less noise.
\label{fig6}}
\end{figure}

\begin{figure}
\epsscale{0.7}
\plotone{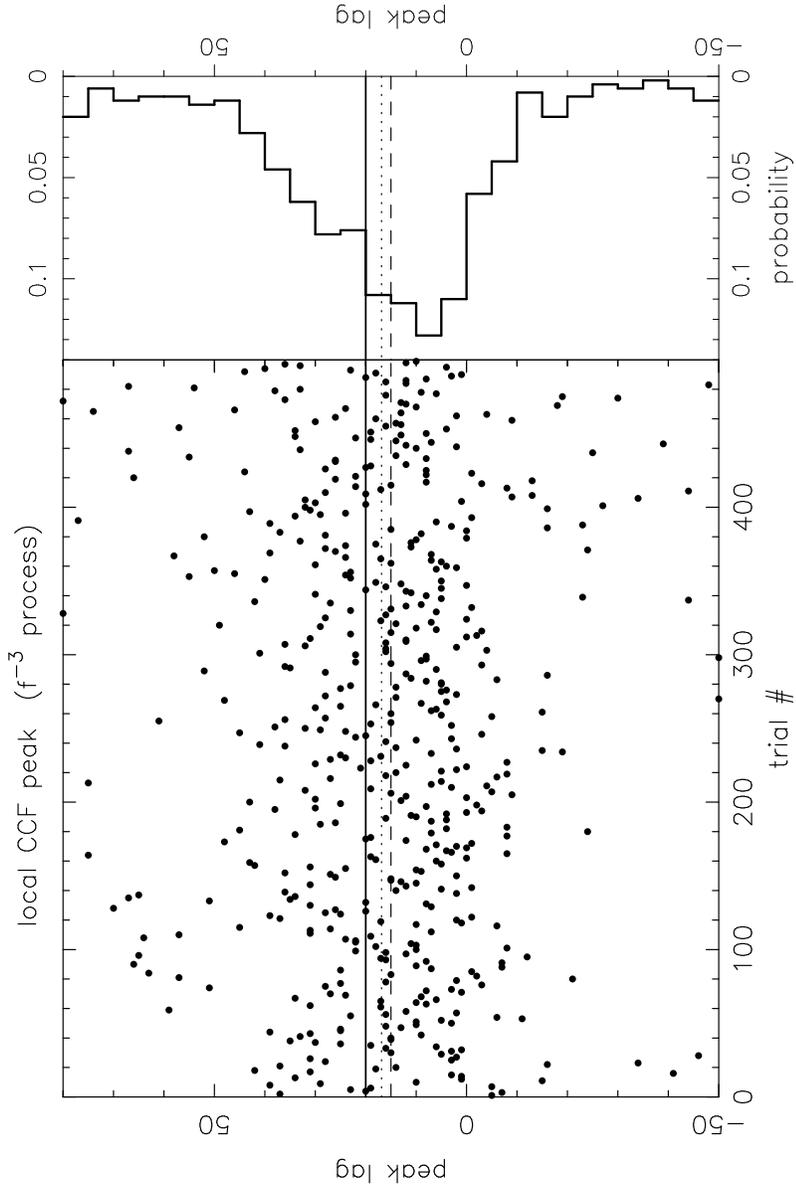}
\caption{The results of 500 measurements of the CCF lag based on
simulations using a $f^{-3}$ power--law power spectrum. Compare with
Figs. 3 and 6, but note the change of scale --- the vertical scale is 4
times larger. The very large scatter in lags is due to the lack of rapid
variations in this ``redder'' continuum light curve, and hence, a very
broad ACF and CCF.
\label{fig7}}
\end{figure}

\begin{figure}
\epsscale{0.7}
\plotone{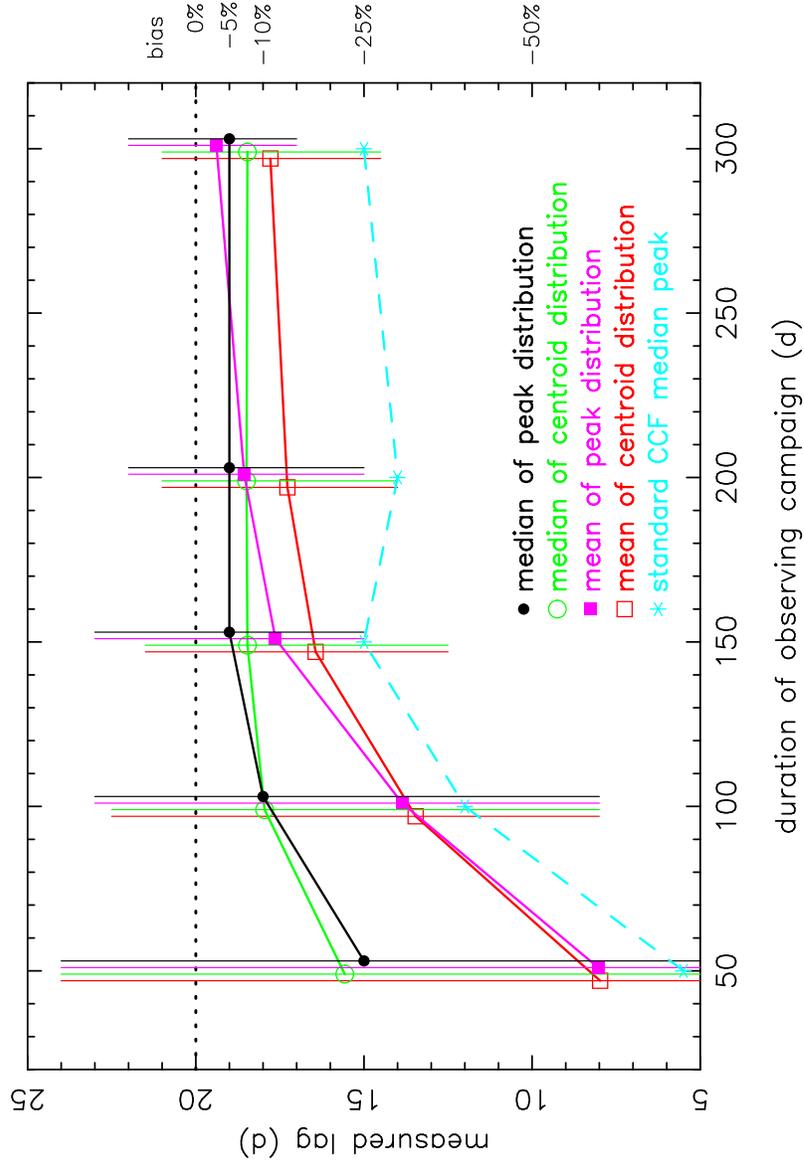}
\caption{
The local CCF lag plotted as a function of the duration of the light
curves.  Shown are the median and mean values of the peak and centroid of
the 1000  simulations. The true lag is 20 days. 
The error bars represent the 68\% limits of the distributions.
As expected, the bias and variance decrease with increasing duration
of the observations. The bias is still $\sim$5\% even with a duration
15 times the lag. The figure also shows that the median values are less
biased than the mean values. For comparison, the median of the peak values
of the standard CCF is also shown. 
\label{fig8}}
\end{figure}

\begin{figure}
\epsscale{0.7}
\plotone{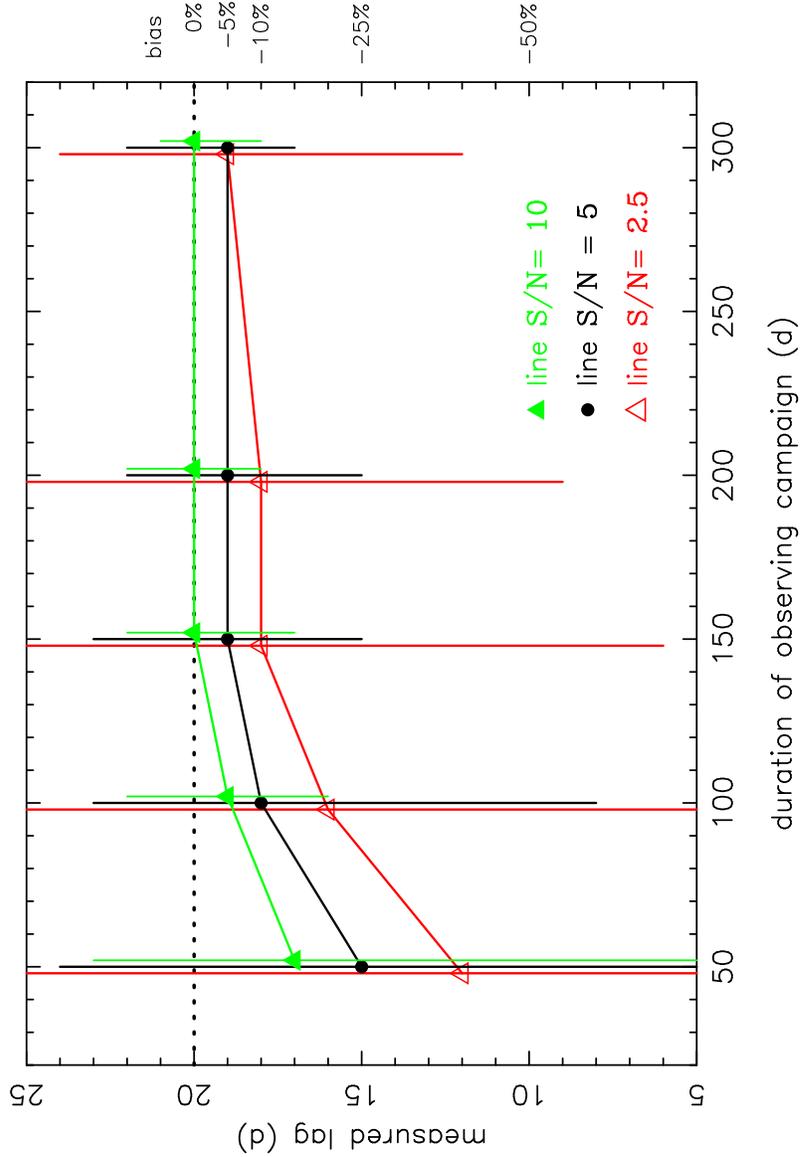}
\caption{The median value of the local CCF lags are plotted against the
duration of the observing campaign for three different values of
signal--to--noise ratio (S/N). The S/N is defined as the intrinsic
RMS variations divided by the observational noise per point for the line
light curve. A S/N of 5 approximates the NGC 5548 H$\beta$ case.
\label{fig9}}
\end{figure}

\begin{figure}
\epsscale{0.7}
\plotone{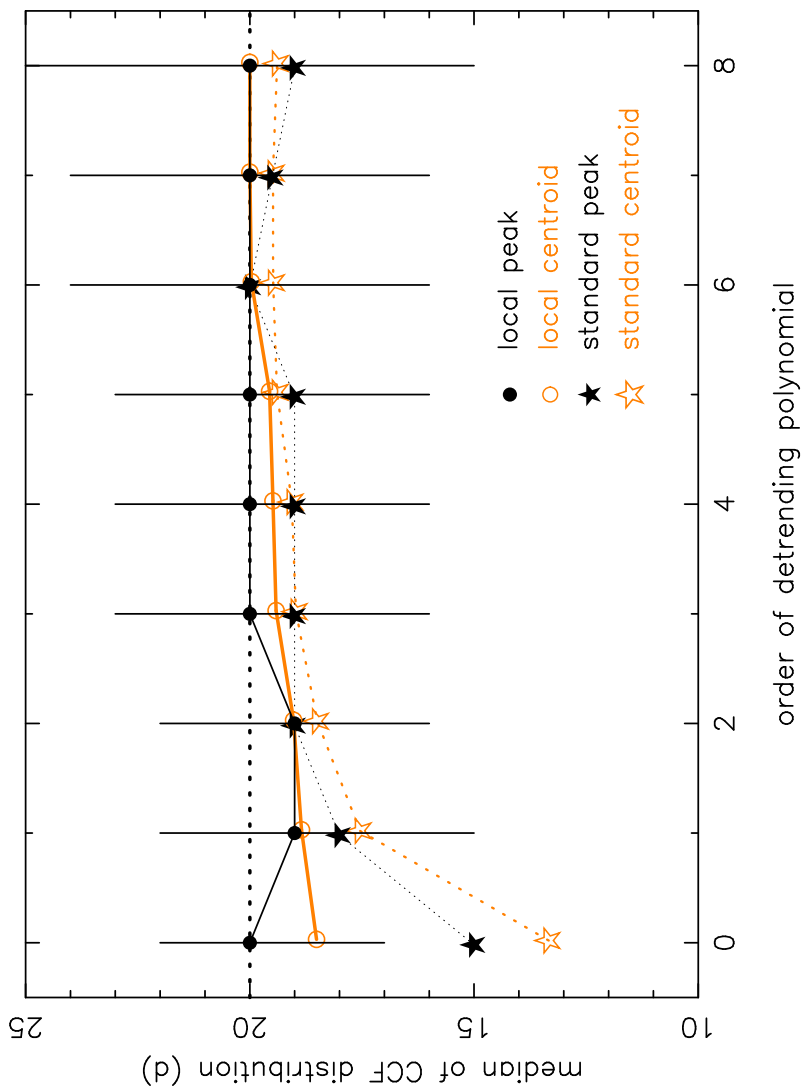}
\caption{The median CCF peak and centroid lags for different amounts of
detrending. For order=0 only the mean has been removed; for order=1
a linear trend is removed; for order=2 a quadratic trend is removed, etc..
The error bars represent the 68\% limits of the local peak CCPD.
Moderate detrending greatly improves the standard CCF estimates, while the
local CCF is not substantially improved. For large amounts of detrending,
the variance in the CCF rapidly grows, negating the reductions in bias.
\label{fig10}}
\end{figure}


\begin{deluxetable}{llclclcl}
\footnotesize
\tablecaption{ Simulation Parameters 
\label{tbl-1}}
\tablewidth{0pt}
\tablehead{ 
\colhead{} & \colhead{} & \colhead{Simulations}  & \colhead{} &
\colhead{} & \colhead{} & \colhead{NGC 5548\tablenotemark{a}} & \colhead{}
\nl
\colhead{} & \colhead{Continuum\tablenotemark{b}} &
\colhead{} & \colhead{H$\beta$\tablenotemark{c}} &
\colhead{} & \colhead{Continuum} &\colhead{} & \colhead{H$\beta$} }
\startdata
mean  flux         & 10.0 && 7.5  && 9.35 $\pm$ 1.86&&7.56 $\pm$ 1.53 \nl
intrinsic rms flux & 2.0  && 1.5  && $\sim$1.82     &&$\sim$1.50 \nl
$F_{var}$          & 20\% && 20\% && 19.5\%         &&19.9\%     \nl
observational noise&0.333 && 0.30 && $\sim$0.33     &&$\sim$0.28 \nl
\enddata

\tablenotetext{a}{derived from Peterson, et al. (1999)}
\tablenotetext{b}{continuum fluxes are in units of 
$10^{-15}$ erg s$^{-1}$ cm$^{2}$ \AA$^{-1}$. }
\tablenotetext{c}{H$\beta$ line fluxes are in units of 
$10^{-13}$ erg s$^{-1}$ cm$^{2}$. }

\end{deluxetable} 

\end{document}